\documentclass[%
reprint,
superscriptaddress,
amsmath,amssymb,
aps,
pra,
]{revtex4-1}
\usepackage{graphicx}
\usepackage{dcolumn}
\usepackage{bm}
\usepackage{braket}
\usepackage{float}
\usepackage{color}

\begin{document}
	\preprint{APS/123-QED}
	\title{Prethermalization and nonreciprocal phonon transport in \\a levitated optomechanical array}
	\author{Shengyan Liu}
	\affiliation{Center for Quantum Information, Institute for Interdisciplinary Information Sciences, \\Tsinghua University, Beijing 100084, China}
	\affiliation{Department of Physics, Tsinghua University, Beijing 100084, China}
	\affiliation{Department of Physics and Astronomy, Purdue University, West Lafayette, Indiana 47907, USA}
	\author{Zhang-qi Yin}
	\email{yinzhangqi@mail.tsinghua.edu.cn}
	\affiliation{Center for Quantum Information, Institute for Interdisciplinary Information Sciences, \\Tsinghua University, Beijing 100084, China}
	\author{Tongcang Li}
	\email{tcli@purdue.edu}
	\affiliation{Department of Physics and Astronomy, Purdue University, West Lafayette, Indiana 47907, USA}
	\affiliation{Purdue Quantum Center, Purdue University, West Lafayette, Indiana 47907, USA}
	\affiliation{School of Electrical and Computer Engineering, Purdue University, West Lafayette, Indiana 47907, USA}
	\affiliation{Birck Nanotechnology Center, Purdue University, West Lafayette, Indiana 47907, USA}
	\date{\today}
	\begin{abstract}
		We consider an array of optically levitated nanospheres in vacuum and  investigate nontrivial phonon  transports   in this system. The levitated nanospheres are coupled by optical binding.
		Key parameters of this system, such as the interaction range, trapping frequencies and mechanical dissipation, are all highly tunable.
		By tuning the spacing between neighboring spheres, the mechanical dissipation and the trapping frequency of each sphere, counter-intuitive phenomena such as prethermalization and nonreciprocal phonon transport can be achieved. Our system  provides a great platform to investigate novel phonon transport and thermal energy transfer.
	\end{abstract}
	\maketitle
	
	\section{Introduction}
	 An optically levitated dielectric particle in vacuum is  a novel optomechanical system that gains lots of attentions recently \cite{Yin2013b,Romero2010,Chang2010,Li2010,li2011millikelvin}. Many recent studies  focus on the center-of-mass motion \cite{Yin2011,li2011millikelvin}, librational motion \cite{shi2013coupling,hoang2016torsional}, rotation \cite{Arita2013laser,kuhn2017optically,Ahn2018,Reimann2018}, nonlinear dynamics \cite{gieseler2013thermal,ricci2017optically,xiao2017bistability,PhysRevLett.117.173602}, and nonequilibrium thermodynamics \cite{Li2010,rondin2017direct,Hoang2018,e20050326,millen2014nanoscale} of a single levitated nanoparticle. Because of its ultra-high mechanical quality factor, this system is an ideal testbed for macroscopic quantum phenomena \cite{Romero2011,Yin2013,Scala2013b,Yin2017,Ma2017}, and can be used for ultra-sensitive measurements of force, torque, mass and acceleration \cite{volpe2006torque,volpe2007brownian,Zhao2014,Ranjit2016,xu2017detecting,Chen2018}.
	 With the development of related technologies, a nanoparticle has been trapped  in high vacuum and its center-of-mass motion has been cooled to a mean thermal phonon number in the order of $10$ \cite{PhysRevLett.116.243601}. The optically levitated nanoparticle could be further cooled to the quantum ground state through cavity sideband cooling, which will enable us to investigate dynamics of nanoparticles in the quantum regime \cite{Yin2011,2013PNAS..11014180K,yoshihiko2018optical}.
	
	 Multiple  dielectric particles can also be levitated simultaneously \cite{Moore:16,Arita:18} and the
	optical fields can generate forces between them. This phenomenon is so-called `optical binding' \cite{dholakia2010colloquim,burns1989optical,Arita:18}, which can induce  interesting self-organization phenomenon \cite{kar2008long,ng2005photonic,lechner2013cavity}. While there have been many studies about optical binding of particles in liquid \cite{dholakia2010colloquim}, systems of optically bond dielectric particles in vacuum are largely unexplored.
 In this paper, we consider an optomechanical array of levitated nanospheres in vacuum,  and study phonon dynamics and energy transport in it.

Controlling the behavior of phonons  is of great research interests in acoustics, nanoscale thermal energy transfer, and material science. There have been many new developments in controlling phonon dynamics, such as phononic crystal  \cite{wu2004surface,yang2004focusing,khelif2006complete}, phononic Josephson junction \cite{xu2017phononic,barzanjeh2016phonon} and acoustic diode \cite{liang2009acoustic,li2004thermal}. Nonreciprocal phonon transport is one of the most important research directions in this field. Various methods, including nonlinearity \cite{hu2006asymmetric,li2004thermal,bender2013observation,fleury2014sound}, parametric modulation \cite{huang2016nonreciprocal}, $\mathcal{PT}$ symmetry \cite{zhu2014pt,lin2011unidirectional,shi2016accessing} and topological design \cite{wang2015topological,xia2017topological,liu2017model}, have been proposed to realize the nonreciprocal transport. Meanwhile, there are some experiments and theories on prethermalization \cite{gong2013prethermalization,neyenhuis2017observation, mohan2005quasi,kottos2011thermalization}, which describes  quasi-stationary states that stay out of equilibrium for very long time.

	Recently, there is a new approach to tailor the phonon dynamics by using optomechanical interactions. Optomechanical structures have been designed to investigate many-body effects \cite{chen2014photon,chang2011slowing,ludwig2013quantum},  achieve topological transport \cite{peano2015topological,brendel2018snowflake} or pseudomagnetic fields \cite{brendel2017pseudomagnetic}. These works mostly based on cavity-based optomechanical resonator arrays. However,  current manufacturing technologies are not sufficient to fabricate cavity-based optomechanical resonator arrays with desired frequency accuracy and coupling. Different from nanofabricated optomechanical resonators with fixed frequencies, the parameters of an optically levitated optomechanical array that we are proposing here are highly tunable. For example, the trapping frequency of each nanosphere can be tuned by changing the power of each trapping laser.
The coupling strength between neighboring nanospheres can  be easily adjusted by using additional optical binding lasers or changing the positions of nanospheres. By tuning these parameters, we can manipulate the transport behavior of phonons in our system to realize nonreciprocal phonon reflection and prethermalization. Our system  provides a versatile platform to investigate novel phonon transport and  energy transfer phenomena.

	This paper is organized as follows: In Sec.II, we introduce the model, derive its classical dynamics and quantum Hamiltonian. In Sec.III, we give some typical parameters, present our calculation of coupling strength, introduce the master equation for correlation matrix and show the general behavior of energy transport. In Sec.IV, we reveal that the prethermalization could happen in our system if the trapping frequencies meet some conditions.  We also investigate the phonon dynamics at different interaction range and trapping frequencies. In Sec.V, we introduce a feasible method to realize nonreciprocal phonon reflection in our system by tuning the phonon energy gain and loss. In the last section Sec.VI, we give a brief conclusion and prospect of this study.

\section{Model}

	\begin{figure}
		\centering
		\includegraphics[width=1.0\linewidth]{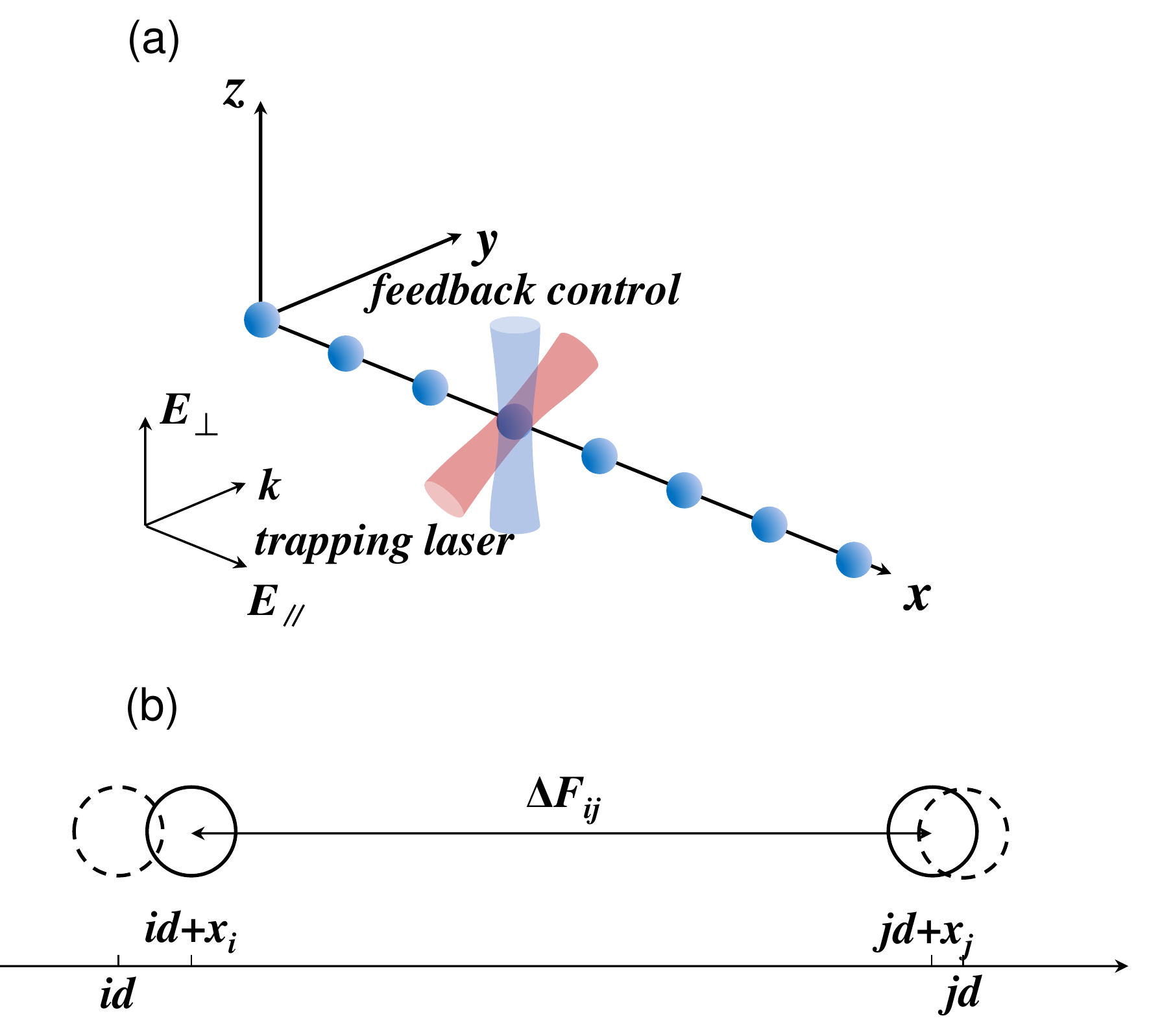}
		\caption{(Color online).(a) Schematic diagram of a 1D levitated optomechanical array. All these nanosphere are placed on $x$-axis and uniformly-spaced. Each nanosphere is trapped by a linearly polarized laser beam propagating along $y$-axis. ${E_\parallel }$ and ${E_\bot }$ denotes the electric field component parallel to the $y$-axis and $z$-axis respectively. (b) The $i$th nanosphere (with equilibrium position $id$) and $j$th nanosphere (with equilibrium position $jd$) deviate their equilibrium positions with displacements of $x_i$ and $x_j$, respectively. $\Delta F_{ij}$ is net force between the two spheres.}
		\label{fig:model}
	\end{figure}
	
	As shown in Fig. \ref{fig:model}, we  consider a levitated nanosphere array formed by identical dielectric nanospheres.  Each sphere is trapped by a linearly polarized Gaussian laser beam. We suppose the diameter of those nanospheres are much less than the wavelength of the trapping beam. In this case, we can use dipole approximation to calculate the force between the nanospheres. For the $i$th nanosphere, the induced dipole ${\mathbf{p_i}} = \alpha {\mathbf{E_i}}$, where $\mathbf{E_i}$ is the total electric field. The total electric field acting on $i$th sphere can be given by the sum of the trapping field $\mathbf{E_i^I}$ and the field generated by other dipole \cite{dapasse1994optical,dholakia2010colloquim},
	\begin{equation}
	\mathbf{E_i} = \mathbf{E_i^I} + \sum\limits_{j \ne i} {{G_{ij}}\mathbf{p_j}} = \mathbf{E_i^I} + \sum\limits_{j \ne i} {\alpha {G_{ij}}\mathbf{E_j}}\label{totalE},
	\end{equation}
	where $G_{jk}$ is the propagator between dipole $j$ and $k$  and $\alpha$ is scalar polarizability of those nanoparticles. We substitute (\ref{totalE}) into itself, we can get
	\begin{equation}
	\mathbf{E_i} = \mathbf{E_i^I} + \sum\limits_{j \ne i} {\alpha {G_{ij}}\left( {\mathbf{E_j^I} + \sum\limits_{k \ne j} {\alpha {G_{jk}}{\mathbf{E_k}}} } \right)}.
	\end{equation}
	As the $\alpha {G_{jk}}$ is small compared with trapping field, we neglect the second order terms. The force can be calculated as $F = \frac{1}{2}{\mathop{\rm Re}\nolimits} \left\langle {{\mathbf{p}^*} \cdot \partial \mathbf{E}} \right\rangle$. So the force acts on the $i$ th sphere reads
	\begin{multline}\label{eq:Force}
	F = \frac{1}{2}{\mathop{\rm Re}\nolimits} \left\langle {{{\left( {\alpha \mathbf{E_i^I}} \right)}^*} \cdot \partial \mathbf{E_i^I}} \right\rangle
	\\ + \frac{1}{2}\sum\limits_{j \ne i} {{\mathop{\rm Re}\nolimits} \left\langle {{{\left( {\alpha \mathbf{E_i^I}} \right)}^*} \cdot \partial \left( {\alpha {G_{ij}}\mathbf{E_j^I}} \right)} \right\rangle }.
	\end{multline}
	The first term is the force due to the trapping laser and the second term is the sum of those optical binding force from other spheres. Eq. \eqref{eq:Force} allows us to calculate the optical binding force ${F_{ij}} = \frac{1}{2}{\mathop{\rm Re}\nolimits} \left\langle {{{\left( {\alpha \mathbf{E_i^I}} \right)}^ * } \cdot \partial \left( {\alpha {G_{ij}}\mathbf{E_j^I}} \right)} \right\rangle $ for each of two spheres first,  and then sum them up. The binding force along the $x$ axis between the two dielectric spheres seperated by $R$ can be written as the sum of two parts \cite{dholakia2010colloquim,samarendra2004optical}
	\begin{equation}
	{F_x} = {F_{xx}} + {F_{xy}},
	\end{equation}
	where
	\begin{equation}
	{F_{xx}} = \frac{{2{\alpha ^2}E_x^2}}{{8\pi {\epsilon _0}{R^4}}}\left[ { - 3\cos kR - 3kR\cos kR + {{\left( {kR} \right)}^2}\cos kR} \right]
	\end{equation}
	and
	\begin{equation}
	{F_{xy}} = \frac{{{\alpha ^2}E_y^2}}{{8\pi {\epsilon _0}{R^4}}}\left[ \begin{array}{l}
	3\cos kR + 3kR\sin kR\\
	- 2{\left( {kR} \right)^2}\cos kR - {\left( {kR} \right)^3}\sin kR
	\end{array} \right].
	\end{equation}
	
	\begin{figure}
		\centering
		\includegraphics[width=1.0\linewidth]{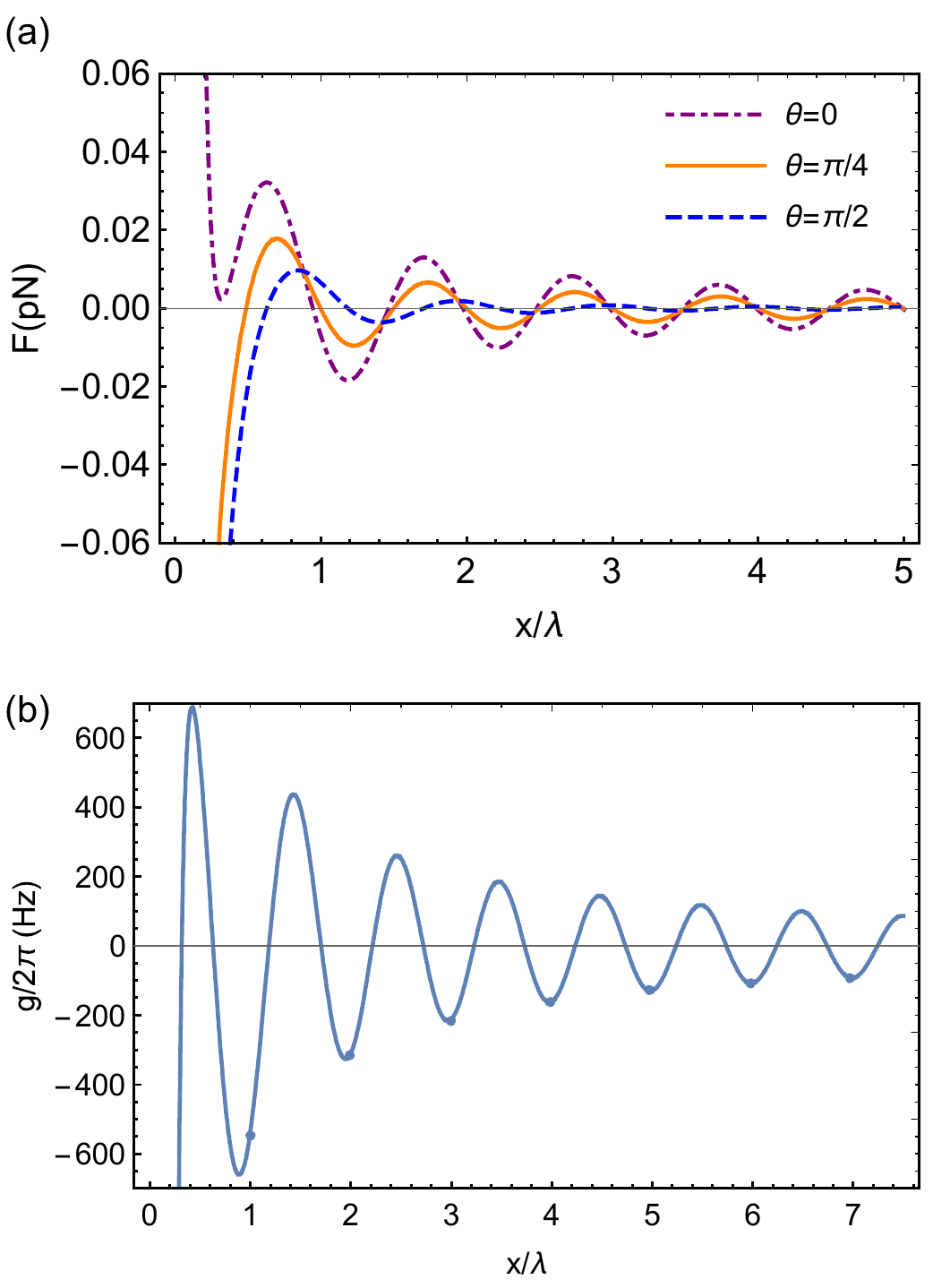}
		\caption{(Color online) (a) The optical binding force between two dielectric spheres as a function of the spacing $x$. $\theta$ is the angle between polarization direction of the trapping laser and the $x$ axis. (b) The coupling strength as a function of spacing $x$ when the polarization direction is perpendicular to the array. Dots represents the values at the spacing we choose.}
		\label{fig:gnew2}
	\end{figure}
	
	Here the two spheres are placed on the $x$-axis and the two trapping beams have the same frequency and polarization. $E_x$ and $E_y$ are the electric field components of the trapping beams parallel to the $x$-axis and the $y$-axis, respectively. Therefore, $F_{xx}$ depicts the binding  force along $x$-axis between $\mathbf{p}_i$ and $\mathbf{p}_j$ due to the $x$ components of the electric field of the trapping lasers, and $F_{xy}$ depicts the binding force along $x$-axis between $\mathbf{p}_i$ and $\mathbf{p}_j$ due to the $y$ electric field components of the trapping lasers.  If the polarization direction is paralleled to $x$-axis, i.e. paralleled to the array, ${E_x} = {E_0}$, ${E_y} = 0$. So ${F_x} = {F_{xx}}$ and in the far field region $kR \gg 1$, ${F_x} \sim {R^{ - 2}}$, which scales similarly to the Coloumb force.  If the polarization direction is perpendicular to $x$-axis, i.e. perpendicular to the array, ${E_x} = 0$, ${E_y} = {E_0}$. So ${F_x} = {F_{xy}}$ and in the far field region $kR \gg 1$, ${F_x} \sim {R^{ - 1}}$, which reduces much slower than the Columb force. These characters allow us to investigate the dynamics under a Columb-like interaction, or under the interaction form that rarely exists in nature. The different force forms are shown in Fig.  \ref{fig:gnew2}(a), where $\theta$ denotes the angle between polarization direction of the lasers and the $x$ axis.
	
	As Fig. \ref{fig:model}(b) shows, we suppose that the uniformly-spaced spheres have small vibrations around their equilibrium positions $x_{i0}$ and focus on the behavior of vibrational energy transport in the array. When the $i$th and $j$th spheres deviate their equilibrium positions with displacement $x_i$ and $x_j$, they will experience a force
	\begin{equation}
	\Delta {F_{ij}} = \Delta R{\left. {\frac{{d{F_x}}}{{dR}}} \right|_{x = \left| {i - j} \right|{d}}} \equiv {k_{ij}}\Delta R.
	\end{equation}
	We denote ${k_n} = {\left. {\frac{{d{F_x}}}{{dR}}} \right|_{R = n{d}}}$ and $\omega_n = \sqrt{k_n/m}$, where $n=\left| {i - j} \right|$. We can write the classical dynamics of the system as
	\begin{equation}
	{{\ddot x}_i} + \omega _0^2{x_i} + \sum\limits_n {\omega _n^2\left( {2{x_i} - {x_{i + n}} - {x_{i - n}}} \right) = 0}.
	\end{equation}
	
	If the nanospheres are cooled to millikelvin temperatures or even to their ground states,  a Hamiltonian in quantum regime is needed to describe the dynamics of the system. The Hamiltonian of the system reads
	\begin{equation}
	H = \sum\limits_i {\left( {\frac{{p_i^2}}{{2m}} + \frac{1}{2}{k_0}x_i^2} \right)} + \sum\limits_{i \ne j} {\frac{1}{2}{k_{ij}}{{\left( {{x_i} - {x_j}} \right)}^2}}.
	\end{equation}
	By using the annihilation(creation) operator ${b_i}({b_i^\dag })$ at site $i$ and neglecting the fast rotating terms \cite{michael2014energy}, we can get
	\begin{equation}\label{eq:Hamiltonian}
	H = \sum\limits_i {\hbar \Omega_i b_i^\dag {b_i}} + \hbar \sum\limits_{i \ne j} {{g_{ij}}\left( {b_i^\dag {b_j} + {b_i}b_j^\dag } \right)},
	\end{equation}
	where
	\begin{equation*}
	\Omega_i = \sqrt {({k_0} + \sum\limits_{j \ne i} {{k_{ij}}} )/m},
	\end{equation*}
	and
	\begin{equation*}
	{g_{ij}} = -\frac{{{k_{ij}}}}{{2m\sqrt{\Omega_i \Omega_j} }}.
	\end{equation*}
	We can cool or amplify the motion of the nanosphere by parametric or force feedback control \cite{PhysRevLett.116.243601,li2011millikelvin}.
	This feedback cooling (amplifying) process can be written in an equation of the average energy $E_i$ of the 	 nanosphere phenomenologically:
	\begin{equation}
	\label{feedback}
	\frac{{d{E_i}}}{{dt}} = - \frac{{{\Gamma ^{cool}}}}{2}(+ \frac{{{\Gamma ^{amp}}}}{2}){E_i}.
	\end{equation}
	We note that it is equivalent to adding $ - i\frac{{{\Gamma ^{cool}}}}{2}b_i^\dag {b_i}$ ($ + \frac{{{\Gamma ^{amp}}}}{2}b_i^\dag {b_i}$) into the Hamiltonian \eqref{eq:Hamiltonian}.

	\section{Master Equation}
	Using the method of  feedback cooling and cavity side-band cooling, we can cool the motion of nanoparticles near their ground states. When we `kick' or drive one of the nanosphere in array, the vibrational energy will quickly spread in the form of local phonons. Here we use the master equation to study the energy transport in the array. The master equation of the system reads
	\begin{equation}
	\dot \rho = - i\left[ {H,\rho } \right] + \sum\limits_j {{\mathcal{L}_j}\rho },
	\end{equation}
	where ${\mathcal{L}_j}\rho$ can be given in Lindblad form,
	\begin{eqnarray*}
		{\mathcal{L}_j}\rho &= {\Gamma _j}\left( {{n_j} + 1} \right)\left( {{b_j}\rho b_j^\dag - \frac{1}{2}\left\{ {b_j^\dag {b_j},\rho } \right\}} \right)\\
		&+ {\Gamma _j}{n_j}\left( {b_j^\dag \rho {b_j} - \frac{1}{2}\left\{ {{b_j}b_j^\dag ,\rho } \right\}} \right),
	\end{eqnarray*}
	where $n_j$ is the average phonon number of the environment around the $j$th nanosphere and $\Gamma_j$ is the mechanical damping rate of the $j$th nanosphere.
	If we define a correlation matrix ${C_{ij}} \equiv \braket{{b_i^\dag {b_j}}}$, and $\braket{\dot C} = Tr({\rho \dot C})$, we use the master equation  and get \cite{asadian2013heat}
	\begin{equation}
	\dot C = i\left[ {W,C} \right] + \left\{ {L,C} \right\} + M,
	\end{equation}
	where ${W_{ij}} = \Omega {\delta _{ij}} + {g_{ij}}$, $L = - \frac{1}{2}Diag\left( {{\Gamma _1},{\Gamma _2}, \ldots ,{\Gamma _N}} \right)$, $M = Diag\left( {{\Gamma _1}{n_1},{\Gamma _2}{n_2}, \ldots ,{\Gamma _N}{n_N}} \right)$. In an experiment, subsequent velocity and energy readout allows us to investigate how energy spreads in the array. Here we suppose that all the spheres have been cooled to millikelvin temperatures and then we excite one of them.  We suppose the system is prepared in a thermal state at beginning
	\begin{equation}
	{\rho _i} = \frac{1}{{\bar n_i + 1}}{\left( {\frac{{\bar n_i}}{{\bar n_i + 1}}} \right)^n}\left| {{n_i}} \right\rangle \left\langle {{n_i}} \right|
	\end{equation}
	and we can get ${C_{ij}} = \braket{b_i^\dag {b_j}} = {\bar n_i}{\delta _{ij}}$, $\bar n_i$ is the initial average phonon number of the $i$th nanosphere. Whether the excitation process is modeled as acting a displacement operator with arbitrary phase on the $i$th sphere or preparing a big fock state $\left| n_i \right\rangle$, we can see that it will only change the corresponding matrix element ${C_{ii}}$. So we might as well set ${\bar n_i} \gg {\bar n_j}\left( {j \ne i} \right)$ and simulate the dynamics of phonon number. Actually, because of ultra-small mechanical loss, the system is very close to an isolated system.
	
	\begin{figure}
		\centering
		\includegraphics[width=1.0\linewidth]{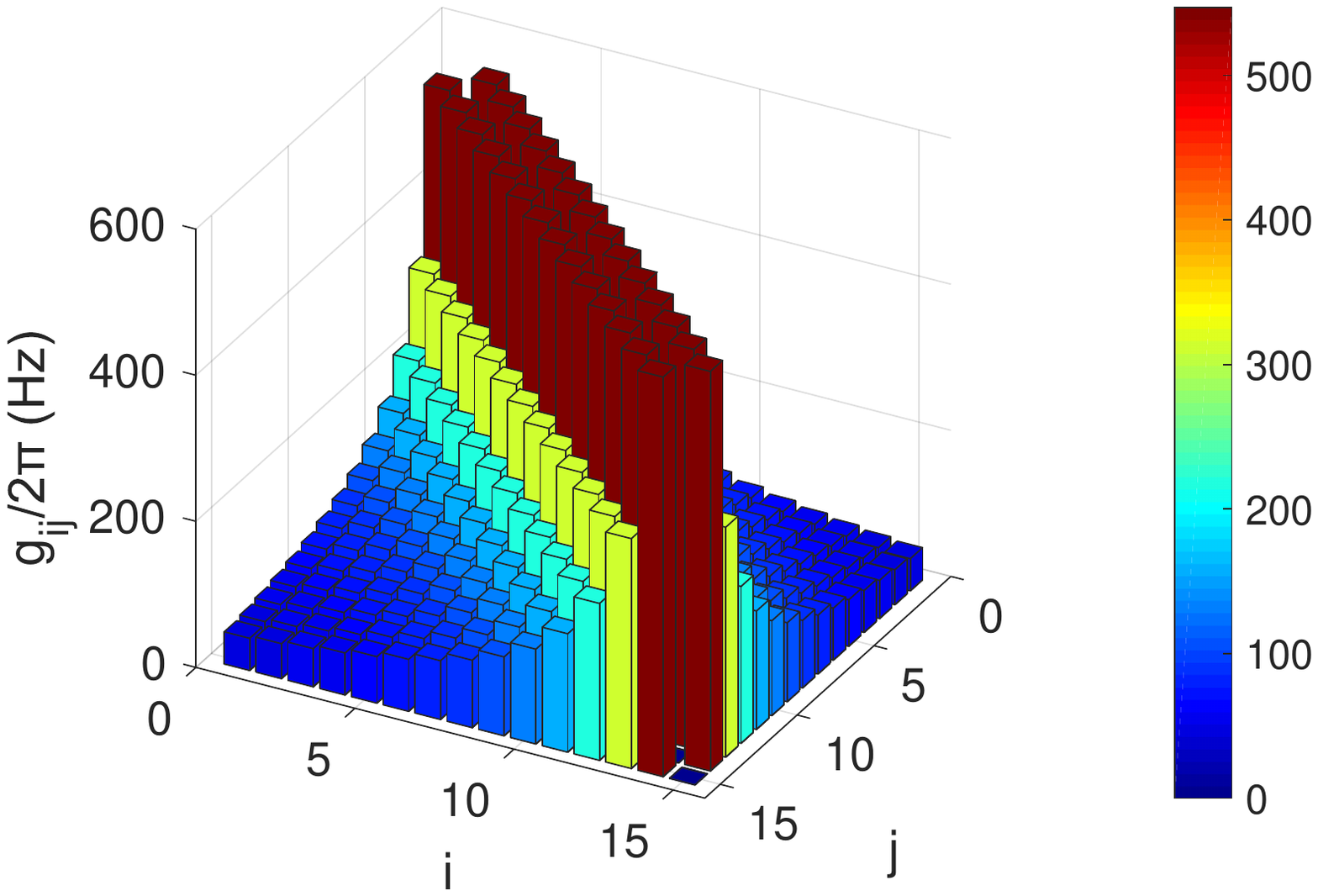}
		\caption{(Color online) The coupling strength between $i$th and $j$th nanospheres.}
		\label{fig:couplingperpendicular15}
	\end{figure}
	
Here we list the parameters we use in this section and Sec. \ref{sec:pretherm}.
	All the nanospheres are made by silica.  The diameter of every nanosphere is 200nm. The power of our trapping laser is $P_0=100$mW, its waist is $w_0=600$nm and its wavelength is $\lambda=1550$nm. Under this situation, we choose the spacing of each two adjacent spheres $d=\lambda$ to achieve a larger coupling. The polarization directions of all laser beam are chosen to be perpendicular to array.  The coupling strengths as a function of $x$ is plotted in Fig. \ref{fig:gnew2}(b),  and the coupling strengths between each two nanospheres under the above parameters are plotted in Fig. \ref{fig:couplingperpendicular15}.
	
\begin{figure}
		\centering
		\includegraphics[width=0.8\linewidth]{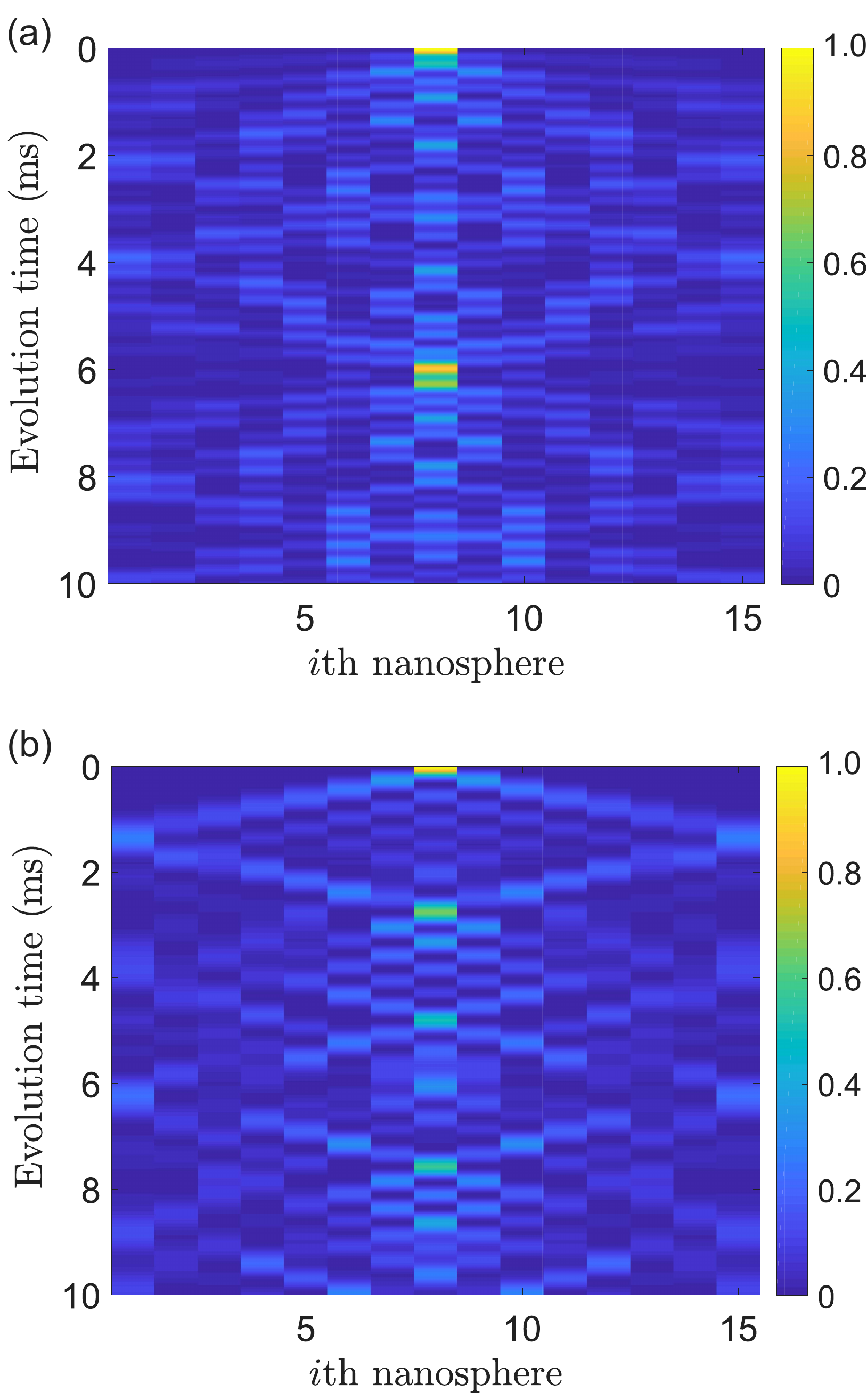}
		\caption{(Color online) Energy diffusion in a levitated optomechanical array with 15 nanospheres. The system is initially prepared in $n_8 \gg {n_j}\left( {j \ne 8} \right)$. (a) Full dynamics with long-range hopping terms due to optical binding. (b) Dynamics with only the nearest-neighbor hopping terms. Energy spreads more uniformly and thoroughly in (a) because of the long-range hopping terms.}
		\label{fig:generalcom}
	\end{figure}

	To have a general picture of phonon transport, the calculation results for the case of $n=15$ and the 8th nanosphere being kicked are shown in Fig.  \ref{fig:generalcom}(a), where  the interactions among the nanospheres are due to long-range optical binding (Fig. \ref{fig:gnew2}(b)).  For comparison, we also calculated a hypothetical case with  only nearest-neighbor coupling, as showed in Fig. \ref{fig:generalcom}(b). It is found that under the long-range optical binding interactions, the energy spreads in the array more uniformly and thoroughly than the nearest-neighbor coupling. We also note that the system can revive because of ultra-small mechanical loss.

	\section{Prethermalization}\label{sec:pretherm}
	How an isolated system reaches thermal equilibrium, and what is its far-from-equilibrium behavior have intrigued researchers for a long time.  Recently, a phenomenon called prethermalization was predicted  theoretically \cite{gong2013prethermalization} and realized experimentally \cite{neyenhuis2017observation} in a trapped ion chain. The
prethemalization is a process that system reaches a quasi-stationary state before it finally thermalizes. In the harmonic oscillator, the prethermalization is also found to happen with nonlinearity \cite{mohan2005quasi,sen2004quasi}.

Since the damping rate of a levitated nanosphere array is extremely small in high vacuum, we can treat our system as an approximately isolated system. We investigate how does the system thermalize when the interaction between trapped nanospheres is long-range interaction or nearest-neighbor interaction.
We find a highly-excited prethermalization phenomenon in our system without nonlinearity by tuning the trapping frequencies. In order to characterize how the system reaches equilibrium, we introduce a parameter \cite{gong2013prethermalization}
	\begin{equation}
	A\left( t \right) = {{\sum\limits_{i = 1}^N {{f_i}{n_i}\left( t \right)} } \mathord{\left/
			{\vphantom {{\sum\limits_{i = 1}^N {{f_i}{n_i}\left( t \right)} } {\sum\limits_{i = 1}^N {{n_i}\left( 0 \right)} }}} \right.
			\kern-\nulldelimiterspace} {\sum\limits_{i = 1}^N {{n_i}\left( 0 \right)} }},
	\end{equation}
	where ${f_i} \equiv {{\left( {2i - N - 1} \right)} \mathord{\left/
			{\vphantom {{\left( {2i - N - 1} \right)} {\left( {N - 1} \right)}}} \right.
			\kern-\nulldelimiterspace} {\left( {N - 1} \right)}}$ and ${n_i} \equiv b
	_i^\dag {b_i}$. The expectation value $\left\langle A \right\rangle \in \left[ { - 1,1} \right]$. It can be easily checked that ${\left\langle A \right\rangle _{t = 0}} = - 1$ for kicking leftmost nanosphere, ${\left\langle A \right\rangle _{t = 0}} = 1$ for kicking rightmost nanosphere and $\left\langle A \right\rangle = 0$ for the system states with bilateral symmetry. So $A$ qualifies the average position of the phonons.
	We also introduce
	\begin{equation}
	\bar A\left( t \right) = \frac{1}{t}\int_0^t {\left\langle {A\left( \tau \right)} \right\rangle d\tau }
	\end{equation}
	to depict the long-time behavior of the array.

The phonon population distribution of quasi-stationary state can be determined by the generalized Gibbs ensemble\cite{neyenhuis2017observation}.
	We can diagonalize $W$ by an orthogonal matrix $U$, such that $UWU^{\dag}=\epsilon_k \delta_{kl}$ and
	\begin{equation}
	{H_{Diag}} = \sum\limits_k {{\epsilon _k}c_k^\dag {c_k}}
	\end{equation}
	where ${c_k} = \sum\limits_l {{U_{kl}}{b_l}}$. The generalized Gibbs ensemble gives the density matrix
	\begin{equation}
	\rho_{GGE} = \frac{{\exp ( { - \sum\limits_k {{\lambda _k}c_k^\dag {c_k}} } )}}{{\mathrm{Tr}( {\exp ( { - \sum\limits_k {{\lambda _k}c_k^\dag {c_k}} } )} )}},
	\end{equation}
	where $\lambda_k$ can be derived by ${\braket{c_k^\dag {c_k}} _{t = 0}} = {\braket{c_k^\dag {c_k}}_{GGE}}$. So that we can get
	\begin{equation}
	\begin{split}
	\braket{b_i^\dag {b_i}}&=\mathrm{Tr}({{\rho _{GEE}}b_i^\dag {b_i}})=\mathrm{Tr}({{\rho _{GEE}}\sum\limits_{k.l} {{U_{ik}}{U_{il}}c_k^\dag {c_l}} })\\
	&= \sum\limits_k {U_{ik}^2\braket{{c_k^\dag {c_k}}}_{GEE}} = \sum\limits_k {U_{ik}^2\braket{{c_k^\dag {c_k}}}_{t=0}}.
	\end{split}
	\end{equation}

	\begin{figure}
		\centering
		\includegraphics[width=0.9\linewidth]{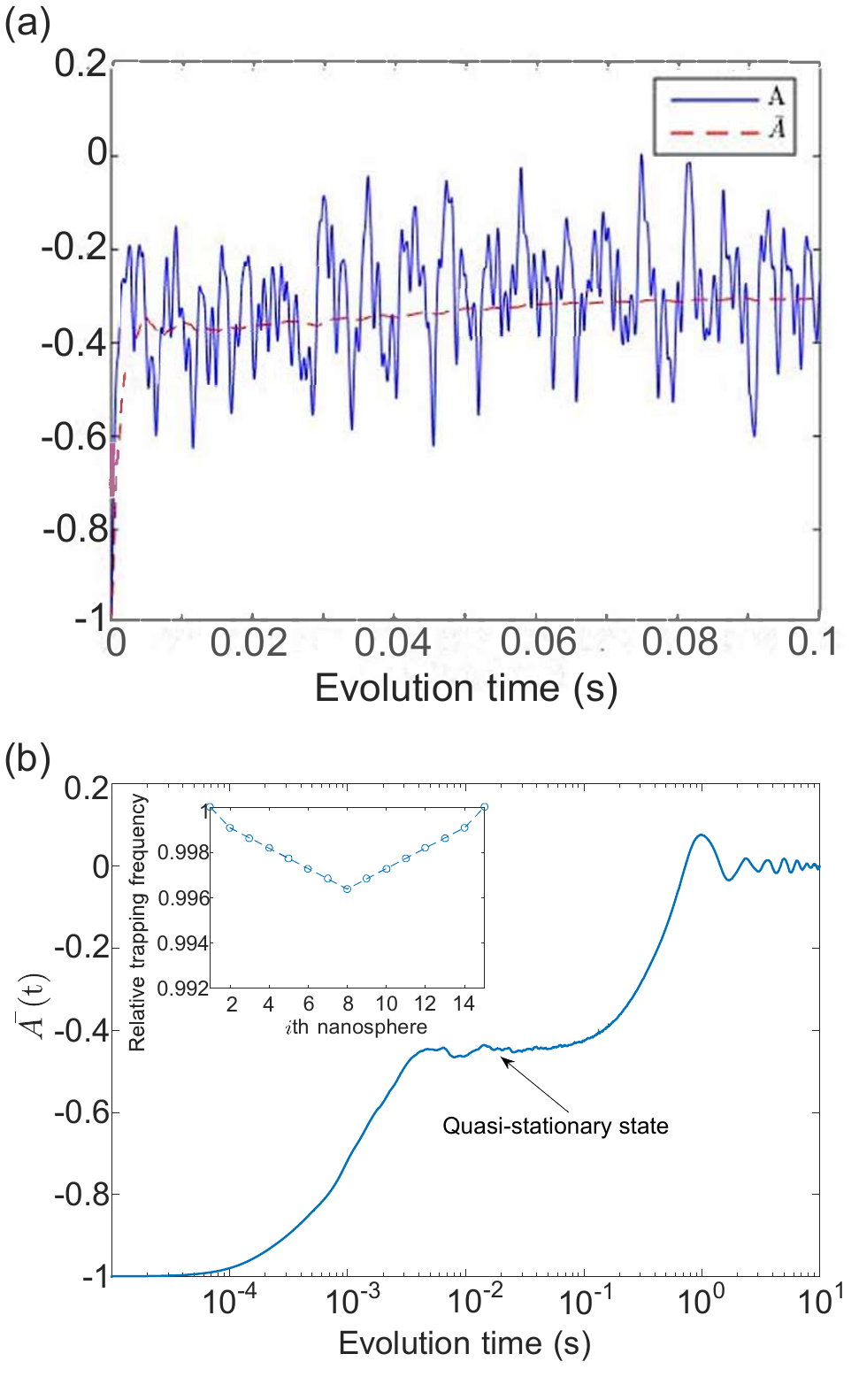}
		\caption{(Color online) (a): Short-time dynamics of $A$ and $\bar{A}$ with long-range interaction. (b): Long-time dynamics of $\bar{A}$ with long-range 	interaction. The inset shows the trapping frequency of each nanosphere. The trapping lasers are all polarized along $z$-axis. So the interaction is roughly proportional to $r^{-1}$ as shown in Fig.  \ref{fig:couplingperpendicular15}.}
		\label{fig:trynew}
	\end{figure}

	\begin{figure}
		\centering
		\includegraphics[width=1.0\linewidth]{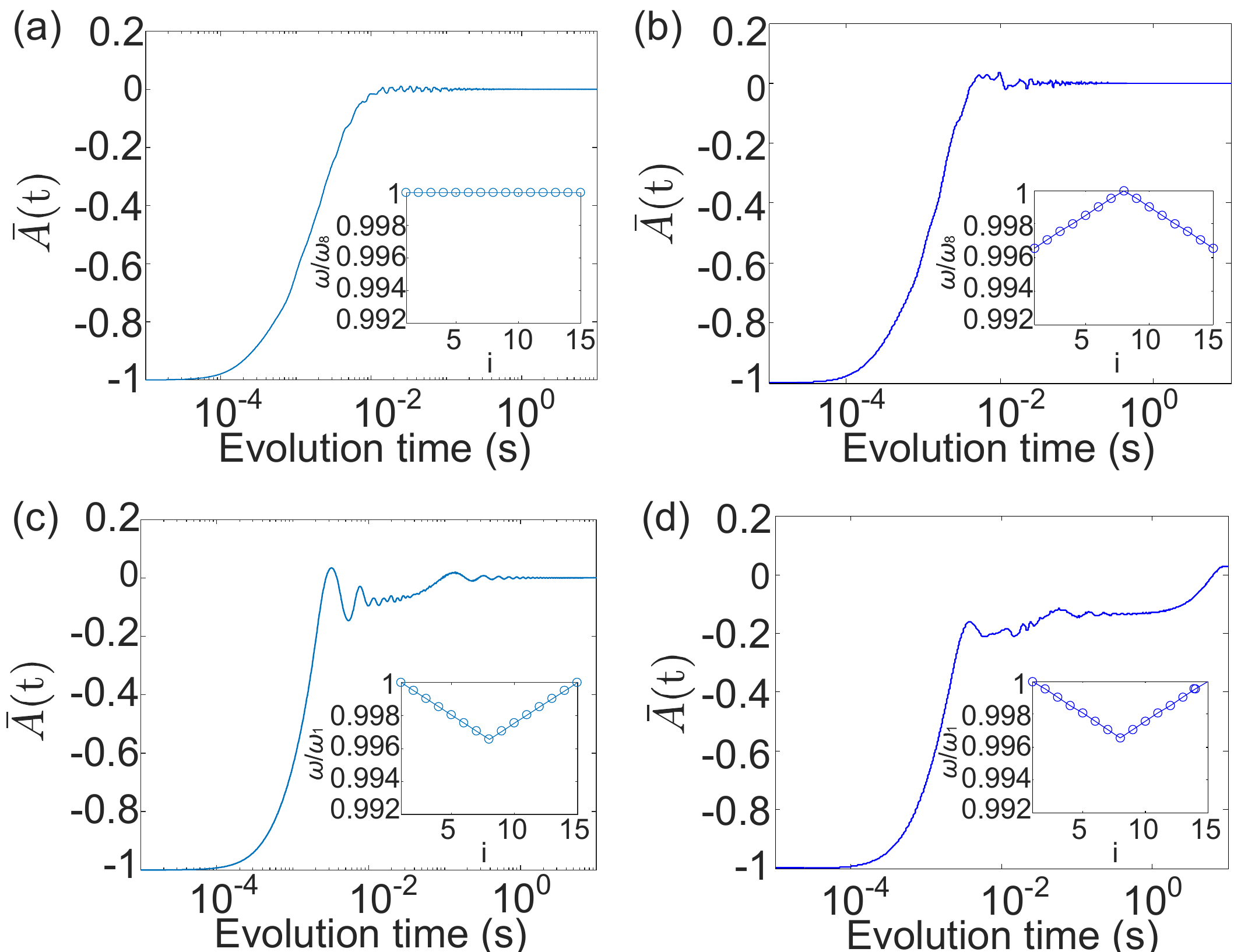}
		\caption{(Color online) Long-time dynamics of $\bar{A}$ under different conditions. The insets show the trapping frequency of each nanosphere. (a): All the trapping frequencies are the same, and $g \sim r^{-1}$ are showed in Fig.  \ref{fig:couplingperpendicular15}. (b): Trapping frequencies of nanospheres near middle are bigger, and $g$s are shown in Fig.  \ref{fig:couplingperpendicular15}. In (a) and (b), prethermalization will not happen. (c) Trapping frequencies are same with Fig. \ref{fig:trynew} with only near-neighborhood hopping terms. (d) Trapping frequencies are same with Fig. \ref{fig:trynew} with coupling strength $g \sim r^{-2}$. In (c) and (d), prethermalization will still happen, but will be less obvious than Fig. \ref{fig:trynew}. }
		\label{fig:final}
	\end{figure}
	
	When the angle between the polarization direction of traps and $x$-axis is $\theta$, we can derive the coupling strength between $i$th and $j$th nanosphere
	\begin{equation}
	{g_{ij}} = {g_{ij, \bot }}{\sin ^2}\theta + {g_{ij, \parallel }}{\cos ^2}\theta,
	\end{equation}
	where ${g_{ij, \bot }}$ and ${g_{ij, \parallel }}$ is the coupling strength when $\theta=\pi/2$ or $\theta=0$. As we discussed in section \uppercase\expandafter{\romannumeral2}, when we turns $\theta$ from 0 to $\pi/2$, the form of the $g$ will change from proportional to $r^{-2}$ to proportional to $r^{-1}$. So, it's easy to change the interaction range by adjusting the polarization direction of the laser beams  to investigate the different behavior of prethermalization.

We assume that the left-most nanosphere is kicked and solve the dynamics of the system under different situations. The interaction is proportional to $r^{-1}$ as shown in Fig. \ref{fig:couplingperpendicular15}. We can find that, by tuning the trapping frequencies properly, for example the trapping frequencies in the middle are smaller, the prethermalization can appear as shown Fig. \ref{fig:trynew} (a) and (b). Before the $A$ (or $\bar A$) increases to 0 which marks thermalization, the system will arrive at a quasi-stationary state with non-zero $A$ and stay there for a long time (Fig.  \ref{fig:trynew} (a)).  If we take all trapping frequencies to be the same or bigger for middle spheres, as shown in Fig. \ref{fig:final} (a) and (b), prethermalization will not happen.
	Fig. \ref{fig:final} (c) and (d) shows the long-time dynamics of $\bar{A}$ with nearest-neighborhood coupling and coupling strength proportional to  $r^{-2}$. As the Fig. \ref{fig:final} shows, although the quasi-stationary states aren't as obvious as Fig. \ref{fig:trynew}, the prethermalization will still happen. These results show that the prethermalization relies on both long-range hopping and trapping frequencies. Because we can set coupling and frequencies respectively, the system give us a chance to investigate their contribution to prethermalization.
	
	\section{Nonreciprocal Phonon Transport}
	Controlling photon/phonon transport, especially nonreciprocal transport has received considerable attention due to its wide application. In  optics, nonreciprocal invisibility can be generated by setting $\mathcal{PT}$-symmetric periodic structures \cite{lin2011unidirectional,feng2013experimental,longhi2010laser}. Here we put up a scheme to realize nonreciprocal phononic invisibility in our discrete levitated nanosphere array by using a gain-loss unit.

	\begin{figure}
		\centering
		\includegraphics[width=1.0\linewidth]{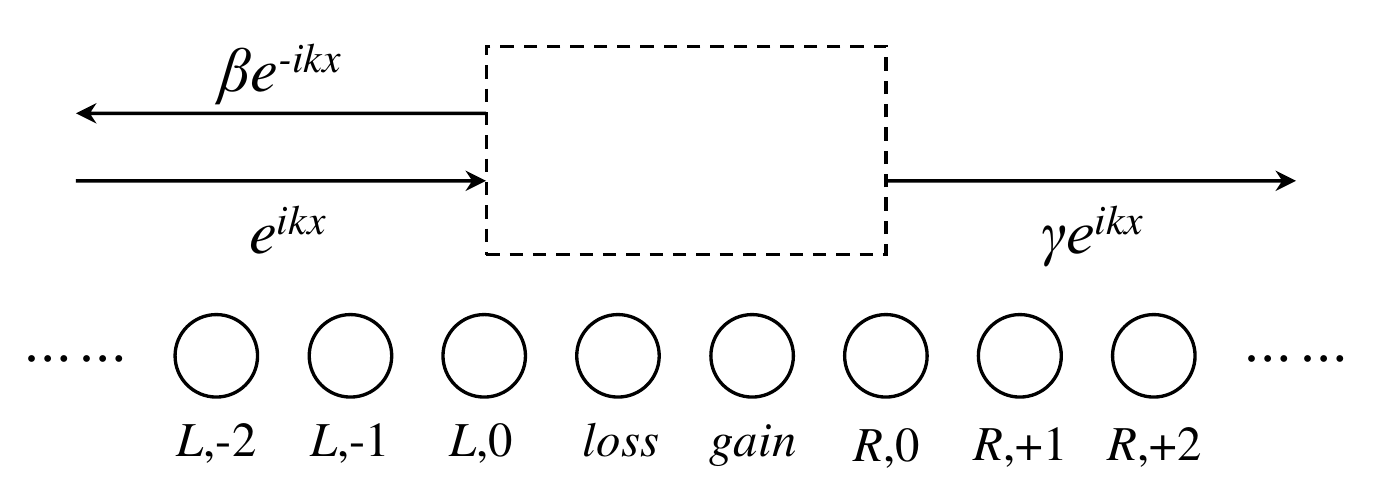}
		\caption{A plane wave incidents from left, scatters and transmits because of the imperfect. Left sphere in the box losses energy while the right gains energy both with rate $\Gamma$.}
		\label{fig:one-way}
	\end{figure}
		
	In last section, we consider the system that is not in an eigenstate of the full coupled set of oscillators. Here, we suppose that the system is in its eigenstate, which means a collective `phonon' transporting as a plane wave in the array. For clarity, at first we only consider the nearest neighborhood coupling in an infinite nanosphere array. Because the mechanical loss is extremely small in high vacuum, so we can neglect intrinsic mechanical loss. We add a gain on a sphere and add a loss on the nearest-neighbor sphere, both with rate $\Gamma$ by using the feedback amplifying (cooling) method in Eq. \eqref{feedback}, as shown in Fig.  \ref{fig:one-way}. The semi-classical evolution functions of the spheres array $L,0$,  $loss$, $gain$, and $R,0$, are
	
	\begin{equation}
	\begin{split}
	E{\alpha _{gain}} &= \left( {\Omega + i\frac{\Gamma }{2}} \right){\alpha _{gain}} + {g}\left( {{\alpha _{R,0}} + {\alpha _{loss}}} \right)\\
	E{\alpha _{loss}} &= \left( {\Omega - i\frac{\Gamma }{2}} \right){\alpha _{loss}} + g\left( {{\alpha _{L,0}} + {\alpha _{gain}}} \right)\\
	E{\alpha _{L,0}} &= \Omega {\alpha _{L,0}} + g\left( {{\alpha _{L, - 1}} + {\alpha _{loss}}} \right)\\
	E{\alpha _{R,0}} &= \Omega {\alpha _{R,0}} + g\left( {{\alpha _{R, + 1}} + {\alpha _{gain}}} \right).
	\end{split}
	\end{equation}
	And the semi-classical evolution fucntions of the spheres far away from the loss-gain unit are in the form of
	\begin{equation}
	E{\alpha _{L(R),i}} = \Omega {\alpha _{L(R),i}} + g({\alpha _{L(R),i - 1}} + {\alpha _{L(R),i + 1}}).
	\end{equation}
	As we can see, the time-reversal symmetry is broken but the system still holds $\mathcal{P}\mathcal{T}$ symmetry with small $\Gamma$. When the gain and loss are big enough, the phase transition will occur and the behavior will be extremely asymmetric.
	
\begin{figure}
		\centering
		\includegraphics[width=1\linewidth]{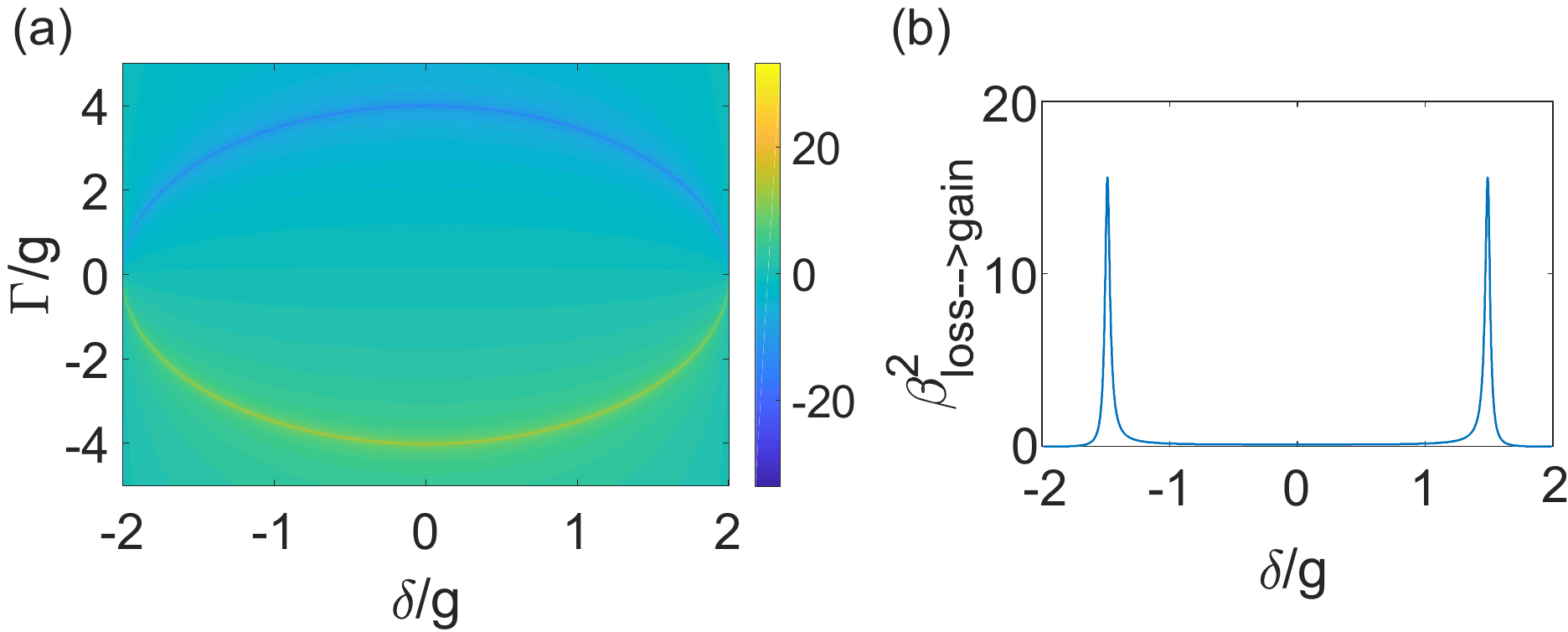}
		\caption{(Color online) (a) $\eta \equiv \ln[(\beta _{gain \to loss}/\beta _{loss \to gain})^2]$ as a function of $\delta/g$ and $\Gamma/g$. Reflection behavior can be asymmetric for some parameters. (b) $\beta_{loss \to gain}$ as a function of $\delta/g$ when $\beta_{gain \to loss}=0$ (i.e. when $\Gamma=2\sqrt{4g^2-\delta^2}$).}
		\label{fig:nnnon}
	\end{figure}

	We take the solution as the form ${\alpha _{L,n}} = {e^{ iknb}} + \beta {e^{-iknb}}$ and ${\alpha _{R,n}} = \gamma {e^{iknb}}$, we can get the transmissivity from left to right
	\begin{widetext}
		\begin{equation}
		{\gamma _{loss \to gain}} = \frac{{8\sqrt {4 - {\delta ^2}} }}{{16i - 2i{\Gamma^2} - 20i{\delta ^2} + i{\Gamma^2}{\delta ^2} + 4i{\delta ^4} - 12\delta \sqrt {4 - {\delta ^2}}  + {\Gamma^2}\delta \sqrt {4 - {\delta ^2}}  + 4{\delta ^3}\sqrt {4 - {\delta ^2}} }}.
		\end{equation}
	The reflectivity is
		\begin{equation}
		{\beta _{loss \to gain}} =  - \frac{{2i({\Gamma^2} + 2\Gamma\sqrt {4 - {\delta ^2}} )}}{{16i - 2i{\Gamma^2} - 20i{\delta ^2} + i{\Gamma^2}{\delta ^2} + 4i{\delta ^4} - 12\delta \sqrt {4 - {\delta ^2}}  + {\Gamma^2}\delta \sqrt {4 - {\delta ^2}}  + 4{\delta ^3}\sqrt {4 - {\delta ^2}} }},
		\end{equation}
	\end{widetext}
	where $\delta= E-\Omega$ and $\delta$ and $\Gamma$ are divided by $g$ for dimensionless. Here we  used the dispersion relation $2{g}\cos kb = E - \Omega $. Take $\Gamma \to -\Gamma$, we can get the transmissivity $\gamma_{gain \to loss}$ and reflectivity $\beta_{gain \to loss}$ for plane wave incidents from the right side, and we can get ${\gamma _{loss \to gain}} = {\gamma _{gain \to loss}}$.
	
	when $\Gamma=0$, the model goes back to the trivial case where $\beta=0$ and $|\gamma|=1$ and the system is reciprocal. When $\Gamma \ne 0$, the system shows nonreciprocal features with $\beta_{gain \to loss} \ne \beta_{loss \to gain}$. We define $\eta \equiv \ln \left[ {{{\left( {{{{\beta _{gain \to loss}}} \mathord{\left/
							{\vphantom {{{\beta _{gain \to loss}}} {{\beta _{loss \to gain}}}}} \right.
							\kern-\nulldelimiterspace} {{\beta _{loss \to gain}}}}} \right)}^2}} \right]$ to quantify the degree of energy transport asymmetric. And as Fig. \ref{fig:nnnon} (a) shows, the energy transport can be extremely asymmetric. At $\Gamma=2\sqrt{4g^2-\delta^2}$ or $\Gamma=-2\sqrt{4g^2-\delta^2}$, this nonreciprocal behavior is most pronounced.
	
\begin{figure}
		\centering
		\includegraphics[width=1\linewidth]{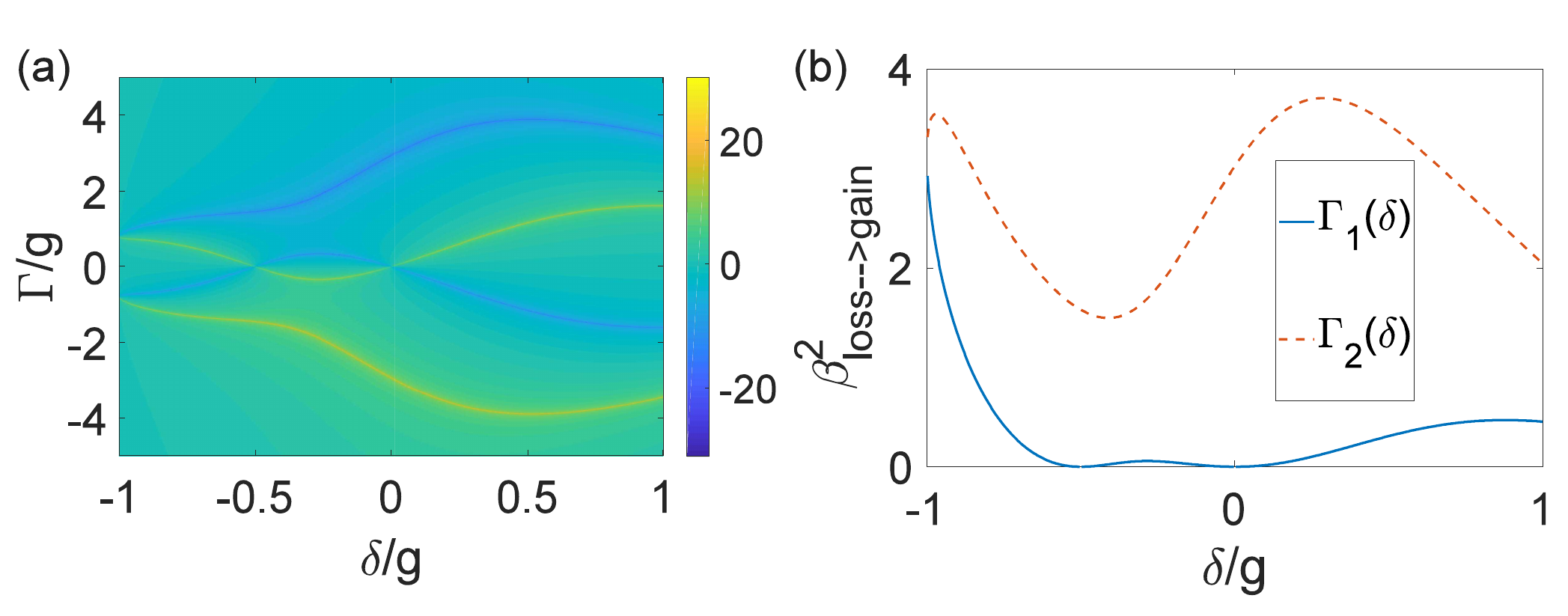}
		\caption{(Color online) Same as Fig. \ref{fig:nnnnon} but considering next-nearest-neighborhood hopping terms. (a) $\eta \equiv \ln[(\beta _{gain \to loss}/\beta _{loss \to gain})^2]$ as a function of $\delta/g$ and $\Gamma/g$. (b) $\beta_{loss \to gain}^2$ as a function of $\delta/g$ when $\beta_{gain \to loss}=0$ (i.e. when $\Gamma=\Gamma_1(\delta/g)$ or $\Gamma=\Gamma_2(\delta/g)$ ) .}
		\label{fig:nnnnon}
	\end{figure}

	For more precise, we can include the next-nearest-neighborhood hopping terms, and the semi-classical equations read
	\begin{equation}
	\begin{split}
	E{\alpha _{gain}} &= \left( {\Omega + i\frac{\Gamma }{2}} \right){\alpha _{gain}}\\
	&+ {g_1}\left( {{\alpha _{R,0}} + {\alpha _{loss}}} \right) + {g_2}\left( {{\alpha _{R, + 1}} + {\alpha _{L,0}}} \right)\\
	E{\alpha _{loss}} &= \left( {\Omega - i\frac{\Gamma }{2}} \right){\alpha _{loss}}\\
	&+ {g_1}\left( {{\alpha _{L,0}} + {\alpha _{gain}}} \right) + {g_2}\left( {{\alpha _{R,0}} + {\alpha _{L, - 1}}} \right)\\
	E{\alpha _{L,0}} &= \Omega {\alpha _{L,0}}\\
	&+ {g_1}\left( {{\alpha _{L, - 1}} + {\alpha _{loss}}} \right) + {g_2}\left( {{\alpha _{gain}} + {\alpha _{L, - 2}}} \right)\\
	E{\alpha _{R,0}} &= \Omega {\alpha _{R,0}}\\
	&+ {g_1}\left( {{\alpha _{R, + 1}} + {\alpha _{gain}}} \right) + {g_2}\left( {{\alpha _{R, + 2}} + {\alpha _{loss}}} \right).
	\end{split}
	\end{equation}
	We take the solution as the form ${\alpha _{L,n}} = {e^{ iknb}} + \beta {e^{-iknb}}$, ${\alpha _{R,n}} = \gamma {e^{iknb}}$ as well and and let $g_2=g_1/2$ approximately. Now, the dispersion relation becomes ${\rm{2}}{g_1}\cos kb + 2{g_2}\cos 2kb = E - \Omega $. We show $\eta=\ln[(\beta_{gain \to loss}/\beta_{loss \to gain})^2]$ in Fig. \ref{fig:nnnnon}. From Fig. \ref{fig:nnnnon} (a), we can find that there are more values of $\Gamma$ and $\delta$ that can realize nonreciprocal reflectivity in the case.
	
	When $\Gamma$ satisfies $\Gamma=\Gamma_1(\delta)$ or $\Gamma=\Gamma_2(\delta)$ where $\Gamma_1(\delta)$ and $\Gamma_2(\delta)$ can be determined numerically, the reflectivity for right-incident wave $\beta_{gain \to loss}=0$ and we show $\beta_{gain \to loss}^2$ in Fig. \ref{fig:nnnnon} (b). As the Fig.  \ref{fig:nnnnon} (b) shows, because of the next-nearest-neighborhood hopping terms, energy can transport without going through the gain-loss unit, so the nonreciprocal effect will decease. We can use multiple gain-loss units to compensate.
	
	\section{Conclusion}
	In summary, we have investigated the dynamics of a levitated optomechanical array. As the system has many adjustable parameters, it is a great platform to study phonon transport behaviors. By tuning the trapping frequency of each nanosphere, we study the phononic energy transport behavior under different types of coupling forms, and find conditions to realize prethermalization. By adding mechanical dissipation and amplification, we propose a method to realize nonreciprocal phonon transport. In future, we anticipate a levitated optomechanical array will enable a lots of new experiments including topological random walk \cite{rudner2009topological,huang2016realizing,chattara2016effects}, $\mathcal{PT}$ symmetry and so on.

	\begin{acknowledgments}
	We acknowledge Shunyu Yao, Yumin Hu and Zhexuan Gong for helpful discussions. Z.Y. is supported by National Natural Science Foundation of China NO. 61771278, 61435007, and the Joint Foundation of Ministry of Education of China (6141A02011604). T.L. is supported by the NSF under Grant No. PHY-1555035 and the Tellabs Foundation.
	\end{acknowledgments}
	
	

%

\end{document}